# Revealing room temperature ferromagnetism in exfoliated Fe$_5$GeTe$_2$ flakes with quantum magnetic imaging

Hang Chen[1]*, Shahidul Asif [2]*, Matthew Whalen[1]*, Jeyson Támara-Isaza[1,3], Brennan Luetke[1], Yang Wang[1], Xinhao Wang[1], Millicent Ayako[1], Saurabh Lamsal[1], Andrew F. May[4], Michael A. McGuire[4], Chitraleema Chakraborty[1,2], John Q. Xiao[1,†], and Mark J.H. Ku[1,2,†]

[1]*Department of Physics and Astronomy, University of Delaware, Newark, Delaware 19716, United States*
[2]*Department of Materials Science and Engineering, University of Delaware, Newark, Delaware 19716, United States*
[3]*Departamento de Física, Universidad Nacional de Colombia, Bogotá D.C., Colombia*
[4]*Materials Science and Technology Division, Oak Ridge National Laboratory, Oak Ridge, Tennessee 37831, United States*

* These authors contributed equally to this work.
† Email: mku@udel.edu, jqx@udel.edu

# Abstract

Van der Waals material Fe$_5$GeTe$_2$, with its long-range ferromagnetic ordering near room temperature, has significant potential to become an enabling platform for implementing novel spintronic and quantum devices. To pave the way for applications, it is crucial to determine the magnetic properties when the thickness of Fe$_5$GeTe$_2$ reaches the few-layers regime. However, this is highly challenging due to the need for a characterization technique that is local, highly sensitive, artifact-free, and operational with minimal fabrication. Prior studies have indicated that Curie temperature $T_C$ can reach up to close to room temperature for exfoliated Fe$_5$GeTe$_2$ flakes, as measured via electrical transport; there is a need to validate these results with a measurement that reveals magnetism more directly. In this work, we investigate the magnetic properties of exfoliated thin flakes of van der Waals magnet Fe$_5$GeTe$_2$ via quantum magnetic imaging technique based on nitrogen vacancy centers in diamond. Through imaging the stray fields, we confirm room-temperature magnetic order in Fe$_5$GeTe$_2$ thin flakes with thickness down to 7 units cell. The stray field patterns and their response to magnetizing fields with different polarities is consistent with previously reported perpendicular easy-axis anisotropy. Furthermore, we perform imaging at different temperatures and determine the Curie temperature of the flakes at ≈300 K. These results provide the basis for realizing a room-temperature monolayer ferromagnet with Fe$_5$GeTe$_2$. This work also demonstrates that the imaging technique enables rapid screening of multiple flakes simultaneously as well as time-resolved imaging for monitoring time-dependent magnetic behaviors, thereby paving the way towards high throughput characterization of potential 2D magnets near room temperature and providing critical insights into the evolution of domain behaviors in 2D magnets due to degradation.

# 1. Introduction

Quasi-two-dimensional (2D) van der Waals (vdW) crystals with long-range ferromagnetic (FM) order have attracted enormous attention in the past few years for their potential applications in the development of next-generation devices [1-5]. Their 2D nature enables high degree of tunability by electrostatic field, pressure, and strain. As they do not require lattice matching, a wide range of heterostructure can be fabricated with other materials to utilize interfacial phenomena or to create new phases of matter. Compared to traditional magnetic materials with 3D structures, their highly tunable nature and readiness for drastic miniaturization presents 2D magnets as promising platforms for implementing a variety of important applications, including new magnetic memory, spintronics devices, and hybrid quantum systems based on magnons. However, the low Curie temperatures ($T_C$) common to these materials impede their applications at room temperature; for example, both $CrGeTe_3$ and $CrI_3$ have $Tc \approx$ 60 K [6,7], while $Fe_3GeTe_2$ has $T_C \approx$ 230 K [8,9] and 130 K [1] for bulk crystal and monolayer, respectively. Recently, $Fe_5GeTe_2$ has been identified as a cleavable material holding intrinsic magnetic order near the room temperature in both bulk and thin flake limit [2,10-13], which nominates it as a suitable candidate for 2D-based devices. In addition, the magnetization has been shown to be tunable by cobalt and nickel substitution, with $T_C$ increasing upon doping and antiferromagnetic behavior is even observed [8,14,15]. So far, magnetization of $Fe_5GeTe_2$ bulk crystal has been characterized via magnetometry, which does not have sufficient sensitivity for exfoliated flakes that are more relevant for devices. For thin flakes, electrical transport measurement has provided indication of near-room temperature $T_C$ [10,13,16]. However, it is a global measurement which conceals the local magnetic information, and probes magnetism indirectly. A local, sensitive detector of magnetic field generated by magnetization will provide the direct validation of room temperature magnetism in exfoliated thin flakes of $Fe_5GeTe_2$.

In this work, we employ quantum magnetic imaging (QMI) technique based on nitrogen vacancy (NV) centers in diamond to observe room-temperature magnetism in exfoliated $Fe_5GeTe_2$ thin flakes down to 21 nm (7 units cell) and investigate their magnetic properties, including $T_C$ and anisotropy. The NV center in diamond realizes a powerful quantum sensor of magnetic field due to its high-sensitivity and high spatial resolution[17,18]. Therefore, NV magnetometry provides an enabling tool for studying novel phenomena in condensed matters and material science. The use of NV centers as magnetic field probe can be implemented as scanning probe microscopy with single NVs [19-21] or optical wide field imaging with NV ensembles [22,23]. The former allows an ultra-high spatial resolution, with ~nm resolution possible, but due to the need for pixel-by-pixel scanning, is limited in its field of view, is time-consuming to measure multiple samples, and is often limited to point-measurement in a sweep (e.g., field or temperature sweep) which may miss crucial information in the evolution of the samples. Wide-field imaging has resolution limited by optical diffraction (resolution 570 nm in our system), but is capable of parallel acquisition of multiples samples, and hence is suitable for rapid characterization, and can access the evolution of samples across a large area during a sweep. Furthermore, optical wide field imaging is significantly simpler to implement compared to NV scanning probe microscopy.

QMI in this work is realized via a bulk diamond crystal with NV ensembles, and the magnetic stray field is measured with continuous-wave optically detected magnetic resonance (cw-ODMR) [24]. Mapping the stray field of the sample allows the analysis of local magnetic properties in sub-micron scale. We first confirm our QMI is capable of detecting stray fields and that such stray fields originate from room-temperature ferromagnetism in exfoliated $Fe_5GeTe_2$ flakes. Then, we conduct both thickness-dependent and temperature-dependent measurements on the flakes; the results reveal the magnetic phase transition and provide insights on anisotropy in $Fe_5GeTe_2$ flakes. Lastly, we demonstrate the utility of our QMI system in characterizing multiple exfoliated flakes in a wide-field view and demonstrate that unprotected flakes exhibit room-temperature magnetism with thickness down to at least 45 nm. We also show the capability of QMI in time-resolved imaging for monitoring time-dependent magnetic behaviors of $Fe_5GeTe_2$ flakes, which provides insight into the evolution of local properties in 2D magnets due to change of materials such as oxidation.

## 2. Experimental setup and methods

Electronic grade diamonds with a {100}-front facet are commercially obtained (Element Six), created using chemical vapor deposition. Diamonds are implanted with nitrogen (Innovion). Implantation density is in the range of ~$10^{12}$-$10^{13}$ cm$^{-2}$, and implantation energy is 6 keV which leads to an average NV depth ~20 nm [25]. The implanted diamonds are subsequently annealed. We fabricate markers on the diamond surface with NVs via lithography and electron-beam evaporation in order to facilitate identification of flakes.

The bulk single crystals of $Fe_{5-x}GeTe_2$ were grown and characterized as discussed in Refs. [10] and [11]. The crystals were quenched from the growth temperature and washed with ethanol and acetone. These quenched, metastable crystals were then cooled to 10K or less in a magnetometer. This cycling to cryogenic temperatures results in a first-order magnetostructural transition near 100K that increases the bulk Curie temperature from ≈ 270 to 310K. This unusual behavior is discussed in detail in Ref. [11] where such crystals are identified as type 'Q-C'. The Fe content has not been controlled during these growths, which are done in the presence of iodine, and the average composition is expected to be near $Fe_{4.7}GeTe_2$. The $Fe_5GeTe_2$ crystals are initially stored in a vacuum desiccator to avoid sample degradation. We exfoliate flakes on the diamond surface with NVs. Exfoliation was performed with a standard medium tack Blue Plastic Film tape (Semiconductor Equipment Corporation P/N 18074). Exfoliation was done quickly in the air, and then the sample was immediately transferred into an electron-beam evaporation chamber for deposition of a 5 nm Pt layer as a protection layer to prevent degradation. We use Pt layer because it is an easily-grown noble metal with high resistance to the oxidization. In addition, a thin Pt layer has a very large spin Hall effect, which can be used in the future to study spin orbit torque effect on $Fe_5GeTe_2$. In this work, we have investigated both flakes protected with Pt layer and flakes without protective layer.

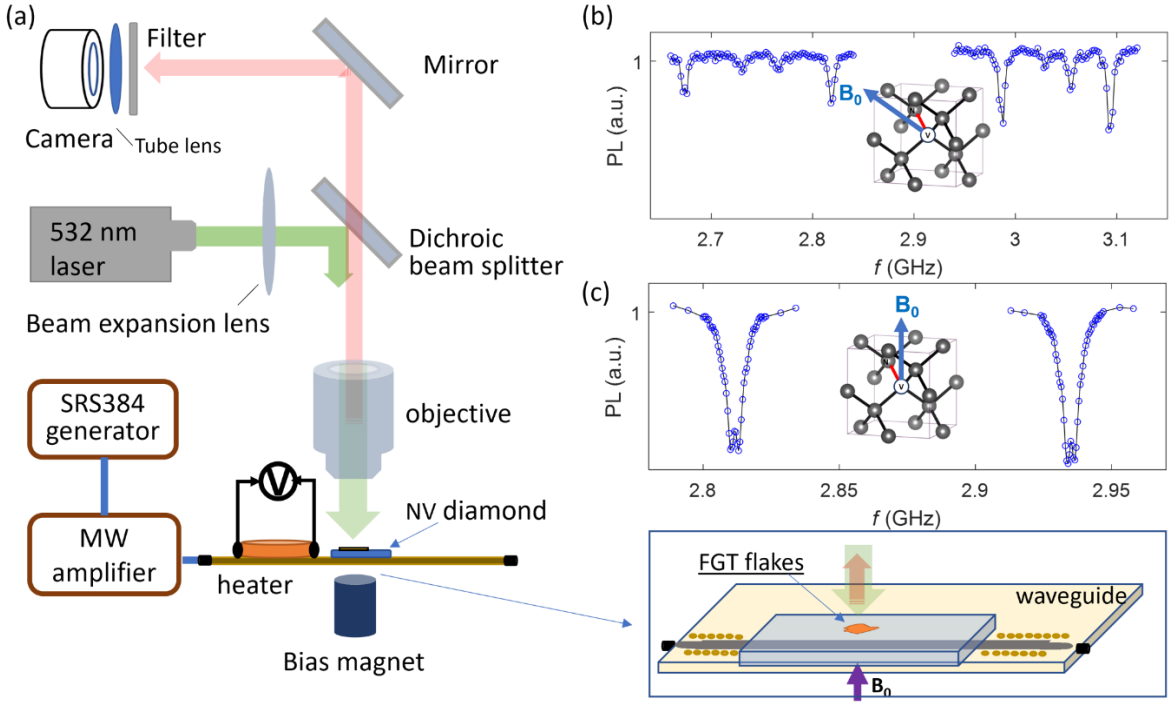

Figure 1. (a) Schematic of the widefield NV QMI. The green laser is directed on Fe$_5$GeTe$_2$ flakes sitting on top of a diamond with NV ensembles near the surface. The PL emitted by NVs (red arrow line) is separated from the excitation via the dichroic beam splitter, further filtered via an interference filter, and imaged onto a camera. The NV diamond is attached on a signal line which delivers microwave for the ODMR measurement. The magnetic bias field is generated by a permanent magnet placed underneath the printed circuit board (PCB). An electrical heater is placed on the PCB near to the flake sample, allowing the temperature dependent measurement on the flakes (b) The bias field is tuned along a direction where one obtains an ODMR spectrum including all 4 pairs of NV spin transitions. (c) The ODMR spectrum measured when the bias field is applied in the z direction.

The schematic of QMI is described in figure 1(a). We use a 532-nm laser (Coherent Verdi 2G) in QMI system. A Kohler-illumination system, consisting of a beam expansion lens and an objective (Olympus ULWD MSPlan80 0.75 NA), expands the laser beam to illuminate an area of about $40 \times 40$ μm$^2$ on the sample. NV photoluminescence (PL) is collected by the same objective. The collected light passes through a 552-nm edge dichroic (Semrock LM01-552-25), after which it is separated from the excitation, passes through another 570-nm long-pass filter to further reduce light not in the PL wavelength range, and is imaged via a tube lens (focal length f = 200 mm) onto a camera (Basler acA1920-155um). Each pixel corresponds to 133 nm on the sample (265 nm if an additional 2x2 binning is applied). Estimated number of NVs that contribute to the signal of a pixel is >10. A SRS384 signal generator supplies microwave (MW) to an amplifier to produce MW with power of 45 dBm. The microwave is delivered into the signal line of a co-planner waveguide on a printed circuit board (PCB), which produces an in-plane MW magnetic field for driving NV spin transition. In our experiment, the diamond surface with exfoliated Fe$_5$GeTe$_2$ flakes and NVs may be either facing up or down.

A permanent magnet, movable via translational stages, supplies a bias field for splitting the ODMR resonances (figures 1(a) and (b)). In principle, there can be up to four pairs of ODMR resonances, each corresponding to $m_s = 0 \leftrightarrow \pm 1$ transitions $f_-$ and $f_+$ of NVs aligned with one of the four diamond crystalline axes; the direction of the nitrogen-vacancy bond provides the quantization axis for the corresponding NV. We employ two configurations of bias field $B_0$. First, bias field can be applied at such a direction as to split all four ODMR resonances (figure 1(b)). In this configuration, we sense $B_{NV}$, projection of the magnetic field along an NV quantization axis of our choice; $B_{NV}$ has contribution from both the out-of-plane and in-plane field. In a second configuration, we align $B_0$ along the $z$ direction (normal to diamond surface), which allows us to sense $B_z$, the projection of field along the $z$ direction. Figures 1(b) and (c) show example ODMR spectra for two configurations of bias field $B_0$. The $B_{NV}$ and $B_z$ can be extracted from ODMR spectra, which allows us to map the stray fields (see **S1** and **S2** sections in Supplemental Material). All measurements are performed in atmosphere. We note that our protocol for measuring stray field rules out artifact due to local strain variation in the diamond, since strain only changes the common-mode shift ($f_{CM} = (f_- + f_+)/2$) instead of differential shift ($\Delta f = (f_+ - f_-)/2$) that we used to extract the stray field (more details are in **S1** section in Supplemental Material).

We perform ODMR by sweeping MW frequency $f$, and at each $f$ we acquire an image, $I(x,y,f)$. To achieve sufficient signal-to-noise ratio (SNR), data acquisition on the order of hours may be required. To deal with sample drift during acquisition, we employ the following innovation in NV wide imaging. We repeatedly perform ODMR measurement to obtain a series of $I_i(x,y,f)$, each of which is saved separately. During post-processing, we align and sum images to obtain a single $I(x,y,f)$. We then fit the ODMR spectrum at each pixel, from which we extract a magnetic field map $B_\alpha(x,y)$, where $\alpha$=NV or $z$ depending on the configuration of bias field employed.

## 3. Results and discussion

We first confirm that NV magnetometer is able to detect the stray field signal from the flakes in ambient condition - at room temperature and in atmosphere, and that the stray field has its origin from ferromagnetism in the flakes. Prior to imaging, $Fe_5GeTe_2$ flakes protected with Pt layer are magnetized by one side (pole 1) of a cylindrical permanent magnet *ex situ* with magnetic field about 0.6 T. We then conduct ODMR measurement with a bias field $B_0 = 40$ G in $z$ direction. A bias field is necessary in order to split the lower and upper ODMR transitions and hence identify the sign of the stray field; the bias field is always applied in the direction of pole 2. The analyzed stray fields of a 100 nm thick flake magnetized by pole 1 is plotted in figure 2(a). Subsequently, we magnetize the flakes again but with the opposite pole (pole 2) of the permanent magnet. We then perform measurement in a similar bias field $B_0 = 30$ G. The field mapping result is shown in figure 2(b). We note that measurements occurred at two different bias fields because the diamond position slightly changes with respect to the permanent magnet underneath the printed circuit board (PCB) when each time the diamond is mounted after the magnetizing procedure. Nevertheless, the signal contribution from the bias field has been removed in all colormaps in figure 2 by subtracting the signal from the area without the flakes, and hence we are able to display Δ$B_z$, the change of stray field from the background. A small and slowly-varying background may remain in Δ$B_z$ due to inhomogeneity of the bias field from the permanent magnet.

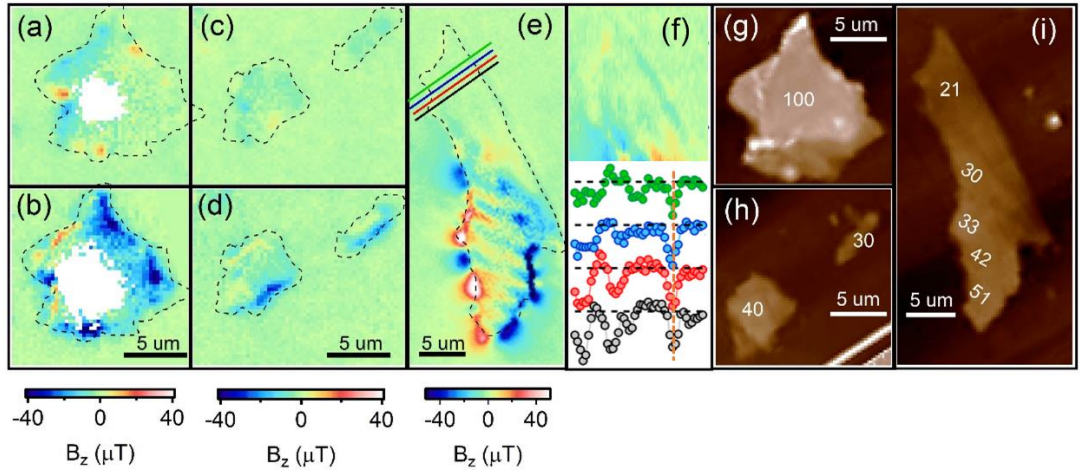

Figure 2. Ferromagnetism in exfoliated thin flakes of $Fe_5GeTe_2$. The NV stray field $B_z$ mappings of the thickest $Fe_5GeTe_2$ flake (center thickness ~100 nm) magnetically initialized by pole 1 in (a) and pole 2 in (b) of a permanent magnet. Flake with thickness of 40 nm (left) and 30 nm (right) magnetized by pole 1 in (c) and pole 2 in (d). These measurements are performed under the $B_z$ configuration. It is noted that data on the pixels with low SNR (e.g., when the error of Δ$B_z$ larger than 3 μT) has been removed in image plotting. (e) $B_z$ stray field mapping of the thinnest flake with thickness of 21 nm. In all images, we display Δ$B_z$, which is the change of $B_z$ from background. (f) Top: zoomed-in view of the upper part of the flake shown in (e), displayed over a range of -20 to +20 μT. Bottom: Stray field variation along the four linecuts marked in (e). Each curve has a vertical offset for clarity of display. Horizontal black dashed lines indicate zero Δ$B_z$ value for each of the curves. Curves are correlated to the linecuts in (e) by different colors. In each curve, the left peak and right dip clearly indicate the boundary of the thinnest part. In particular, a sharp dip manifests along the right boundary, which is marked with vertical orange dashed line. (g)-(i) AFM images of the 4 measured flakes where the thicknesses of flakes are labeled by the numbers in the unit of nanometers. The outlines of the flakes in panel (a) to (e) have been depicted with black dash lines.

A single domain magnet with homogeneous out-of-plane magnetization produces a Δ$B_z$ pattern that has strong amplitude at the flake boundary, while in the inner parts of the flake the stray field is weaker but still significant (see **S3** in Supplemental Material). If the magnetization is in-plane, Δ$B_z$ instead is localized at a pair of opposite edges, and is vanishingly small in the inner part of the flake. More generally, the flakes may have domains or may have magnetic textures, which will produce more complex Δ$B_z$ pattern, hence the experimentally imaged stray field will not necessarily have all the matching features compared to that of an idealized single domain magnet. Nevertheless, in general, one anticipates stray field signal along the boundary and more generally across the flake, if magnetization is present. Indeed, in figures 2(a-e), significant stray field is observed that is associated with the flake. The question then is whether the stray field comes from paramagnetism or ferromagnetically ordered magnetization. If the flakes are paramagnetic, we would expect a stray field that is not affected by the magnetization procedure, and in fact should be slightly weaker in figure 2(b) (with bias field 30 G) compared to in figure 2(a) (bias field 40 G). Instead, we observe the contrary, where figure 2(b) shows a generally much stronger stray field amplitude compared to figure 2(a). Clearly, the measured stray fields show significant difference in strength when we magnetize the flake in opposite directions. This result is consistent with the presence of ferromagnetically ordered magnetization: when the magnetization is polarized along the opposite directions, one polarization will show a stronger magnetization (and hence stray field) in the same or even a smaller bias field. We note that this analysis does not require the flake to have a single magnetic domain, as we have access to local stray field. Our result indicates that the detected signal indeed comes from the magnetization change in the flake. Similar observations are found in other two smaller flakes with thickness of 40 nm and 30 nm as shown in figures 2(c) and (d). Therefore, we conclude the stray field observed is generated by ferromagnetism in the flakes.

The topology of the stray field provides insights in the magnetic anisotropy of the flakes. In figure 2(b), we observe stray field of mostly the same sign (namely negative) along the edge as well as in the inner part of the flake; additional features in the stray field are likely due to domain or magnetic textures. This topology appears to match with the expected stray field distribution induced by a perpendicular magnetization [26]; on the other hand, an in-plane magnetization would have led to $\Delta B_z$ pattern localized to the boundary and is vanishingly small in the inner part of the flake, where in our experiment, stray field that is readily observable is generally present well into the interior of the flake. We note that the center the thicker flakes has low signal-to-noise ratio (SNR). This is because the thick flakes attenuate light and reduces the number of collected photons. When the thickness of the exfoliated flake decreases to 40 nm (left flake in figures 2(c) and (d)), we are able to obtain higher SNR around the flake center. In figure 2(d), we once again note that stray field is generally present across the entire flake. Lastly, we note the contrast between figures 2(a) and (b) (as well as between figures 2(c) and (d)) indicates the change on magnetization induced by a perpendicular magnetizing field persists after the magnetizing field has been removed. Previously, perpendicular anisotropy has been demonstrated in $Fe_5GeTe_2$ bulk crystals via magnetometry, and in flakes via reflective magnetic circular dichroism (RMCD) and anomalous Hall effect (AHE) measurements [10]. Aspects of our results point to easy-axis anisotropy near $T_C$, consistent with prior literatures, though further investigation is required to draw certain conclusions. For instance, the stray field appears to remain across a flake, though the existence of domains could lead to apparent sign inversions within a given flake that may also give the appearance of in-plane anisotropy. Future measurements at higher and variable bias fields would provide even more insight into the magnetism in these and similar materials in this thin flake form, and in particular allow more definitive conclusions about anisotropy.

We then explore what is the thinnest flake we have observed that exhibits room temperature ferromagnetism. We display the stray field map of a 21 nm thick flake in figure 2(e). Figure 2(f) shows a zoomed-in view of the upper part of the flake, which is the thinnest part of this flake; it clearly displays stray field near the boundary (stray fields along four linecuts in figure 2(e) are displayed in figure 2(f)), which demonstrates the ability of our wide-field NV magnetometer to characterize the stray field induced by very thin flake. Hence, we demonstrate $Fe_5GeTe_2$, when protected with a thin Pt layer, exhibits ferromagnetism in flakes as thin as 21 nm, corresponding to 7 unit cells. The thicknesses of all those flakes are measured by the atomic force microscopy (AFM) as shown in figures 2(g-i). We note that in figure 2(i) multiple terraces are present in the bottom half of the flake, as revealed by AFM, and may be responsible for the complex stray field pattern in the bottom half of figure 2(e).

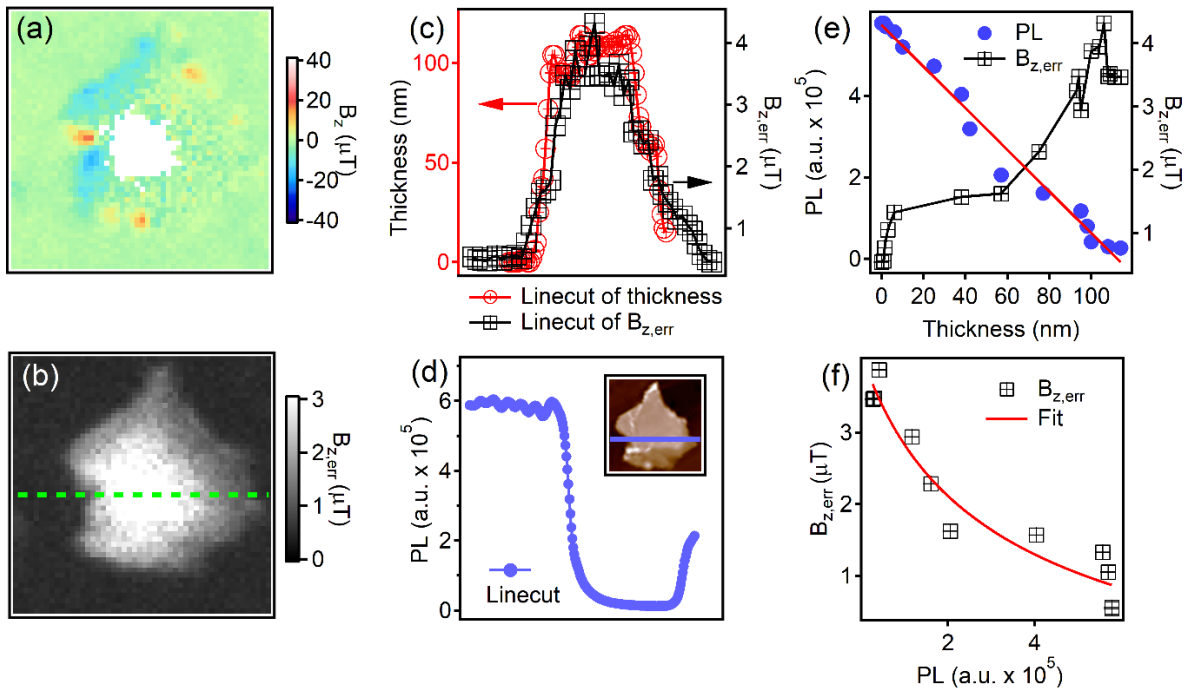

Figure 3. (a) Stray field $B_z$ map of the thickest $Fe_5GeTe_2$ flake. It is the same as figure 2(a). (b) The distribution of stray field uncertainty $B_{z,err}$. (c) Variation of flake thickness and $B_{z,err}$ along the linecuts in (b). (d) Change of PL intensity along the blue linecut in the AFM image of the flake shown in the inset; this corresponds to the same linecut in panel (b). (e) PL intensity and $B_{z,err}$ as a function of flake thicknesses. The data points are extracted from the data in (c) and (d). PL vs thickness is well-described with a linear fit. (f) The relationship between PL intensity and $B_{z,err}$. The data is well described by a fit of the form $B_{z,err} \sim 1/\sqrt{PL}$.

We have noted a variation in SNR across the flake due to thickness-dependent attenuation of PL. The uncertainty $B_{z,err}$ in magnetic field measured with ODMR scales as $\sim 1/N^{1/2}$, where $N$ is the number of photons collected. This indicates that the more photons collected, the lower the $B_{z,err}$.

We directly extract $B_{z,err}$ from our fitting to the ODMR peak using the Lorentzian function. To address the relationship between $B_{z,err}$ and $N$, we take the data from the thickest flake (corresponding to the flake shown in figure 2(a,b)) as an example. Figure 3(a) and (b) show mapping of $B_z$ and $B_{z,err}$, respectively. The pixels we removed in figure 3(a) correspond to areas where $B_{z,err} > 3$ μT, which match the areas with high $B_{z,err}$ in figure 3(b). In figure 3(c), we observe that $B_{z,err}$ along the green linecut in figure 3(b) shows a similar trend as the thickness variation. The photoluminescence (PL) variation along the same linecut is shown in figure 3(d). It is found that $B_{z,err}$ increases, while PL decreases, with flake thickness as shown in figure 3(e). Attenuation of PL due to thickness is well-described by a linear decay. In figure 3(g), the relationship between $B_{z,err}$ and PL can be well-described by a fit to the function $B_{z,err} \sim 1/\sqrt{PL}$, as PL $\propto N$. Hence, we attribute the low SNR (corresponding to large $B_{z,err}$) in the center of the thicker flakes to the increase of light attenuation with flake thickness. Light absorption by 2D materials has been recognized and described in the literatures that attenuation is nearly linear for small layer numbers (≲10), and becomes nonlinear beyond that, but nevertheless increases monotonically with increasing layer numbers[27-30].

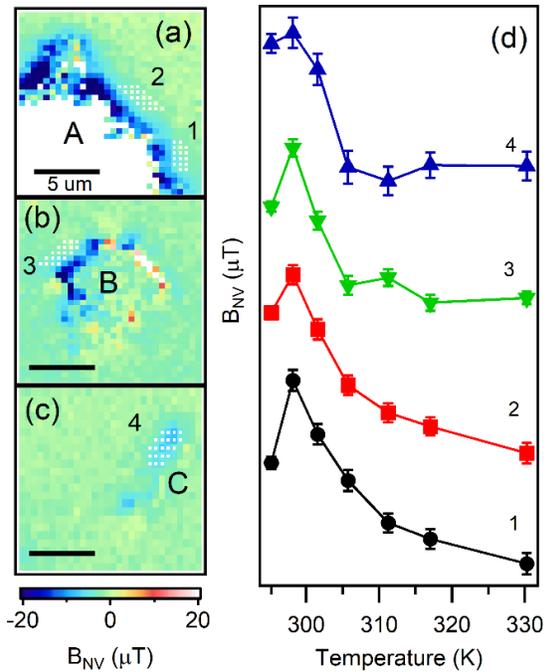

Figure 4. (a)-(c) NV stray field images $B_{NV}$ of $Fe_5GeTe_2$ flakes A, B and C at room temperature. The numbers label the areas where we investigate the temperature dependence of stray field. The pixel points of these areas are marked by the white dots. (d) The temperature dependence of stray field averaged from the selected area. Each curve is offset in the y-axis for clarity. The black circles, red squares, green down-pointing triangles and blue up-pointing triangles indicate the averaged stay fields of area 1, 2, 3 and 4, respectively.

Having confirmed that the flakes are ferromagnetic at room temperature, a natural question to ask is what are the Curie temperatures of the flakes. To answer this question, we take advantage of the ability to perform simultaneous magnetometry and thermometry with NVs to map the stray field at different temperatures. In this measurement, an electrical heater is placed on the PCB near the diamond chip to enable temperature tuning above the room temperature. Here, we employ the configuration in figure 1(b) to measure $B_{NV}$, which is more convenient for simultaneous magnetometry and thermometry.

We first heat up the sample to 330 K which is higher than the widely reported Curie temperature $T_C$ of $Fe_5GeTe_2$ material [2,10,13], and then perform magnetic imaging measurements as we continuously decrease the temperature down to room temperature. This allows us to form a magnetic phase diagram of the magnetization by examining the magnetization-induced stray field, from which the $T_C$ can be determined. At several temperatures, we perform ODMR measurement to obtain magnetic images. The same ODMR measurement allows us to extract the temperature (see **S4** in Supplemental Material). In this work, temperature fluctuation is less than 1 K within a measurement cycle at each temperature point. We examine the temperature dependence of $B_{NV}$ of the 3 flakes in figures 2(g) and 2(h). In figure 4(a), we show the stray field mapping measured at room temperature and mark the areas where we focus on the investigation of temperature dependence. For the thicker flakes A (the same flake as in figures 2(a) and (b)) and B (the flake on the left of figures 2(c) and (d)), we necessarily have to focus on the stray field closer to the edges due to the lower SNR in the center, as noted before, where for the thinner flake C we are able to look at the stray field right at the flake. We note that here we measure $\Delta B_{NV}$, which has contribution from both out-of-plane and in-plane stray field, and hence has different topology compared to the measurement in figure 2. This configuration is employed here as it is more convenient for temperature-dependent measurement (see **S5** in Supplemental Material), and we are only concerned with how the local stray field evolves with temperature.

The average stray field of the selected regions are plotted as a function of temperature in figure 4(b). With increasing the temperature, all $B_{NV}$ vs. $T$ curves show an initial increase or no change from room temperature (295 K) to ~300 K, and then quickly decrease to a constant value. We attribute the divergence-like behavior of the curves near ~300 K to the phase transition from ferromagnetic (FM) to paramagnetic (PM) order in $Fe_5GeTe_2$ flakes. The observed initial upturn in each curve has also been observed in $M$ vs $T$ measurement of $CrI_3$, an easy-axis van der Waals magnet, when the bias is in-plane, and is attributed to the softening of anisotropy near $T_C$ [31]. This scenario is consistent with our evidence for out-of-plane anisotropy and the fact that $B_0$ has a significant in-plane component in this measurement. The $T_C$ is determined from the location where the $B_{NV}$ vs. $T$ curve begins a rapid descent, and we find $T_C \approx 300$ K for all selected regions. We do not observe any particular thickness dependence of $T_C$ from figure 4(c), which is consistent with the observation in Ref [10].

In comparison to previous works on exfoliated flakes of the same or similar materials with near-room temperature Curie temperature, $T_c$ =280 K in $Fe_5GeTe_2$ flakes was extracted from AHE measurement [10] and $T_c$ =270 K was obtained in $Fe_4GeTe_2$ flakes with thickness down to 7 units cells from magnetic circular dichroism (MCD) measurement with a large applied external field of 0.5 T [32]. In the present work, we report $T_c$ right at 300 K. Furthermore, we note that it is important to rule out paramagnetism when the temperature is above $T_c$, which may generally contribute to detected signal with a given magnet-characterization technique, especially at large external field. In our work, we are able to observe stray field at a low bias field of a few tens of Gauss. Furthermore, we rule out paramagnetism contribution by initially magnetizing (with a perpendicular magnetic field at 0.6 T) the flake in two opposite directions then imaging the flakes at small bias fields (tens of Gauss) applied in the same direction. We observed a significant difference in the strength of the stray fields, which even switch sign at certain locations, as shown in figure 2(a) and (b). This is not the case if paramagnetism dominates. The magnetization of a paramagnetic material should remain the same in a given external field no matter how one magnetizes the flake; in another word, one should observe identical stray field map in the same bias field, no matter how it was initially magnetized, but we observed the contrary, which rules out paramagnetism. We attribute this opposite sign of the stray fields to the magnetic hysteresis of the flake, and hence we are able to report ferromagnetism in $Fe_5GeTe_2$ flakes at room temperature. In addition, the observed hysteresis loop in isothermal magnetization ($M$~$H$) curve measured on our bulk $Fe_5GeTe_2$ crystal at 300 K (see **S6** in Supplemental Material) also confirms room-temperature ferromagnetism. Therefore, our work has an additional novelty in that we unambiguously confirmed room-temperature ferromagnetism in a thin iron-rich 2D magnet.

Additionally, we also point out a unique feature of NVs as magnetic-characterization tool compared to optical techniques such as MCD, RMCD, or magneto-optic Kerr effect (MOKE). It is known that RMCD/MCD is only sensitive to the out-of-plane component of the magnetization, and hence is not suitable for in-plane magnets. MOKE is sensitive to either in-plane or out-of-plane magnetization, depending on the given configuration between the probing light and magnetization, but not to both simultaneously. Thus, to obtain a signal requires using the correct configuration that corresponds to the magnetization direction, which may not be *a priori* known. Though vector-MOKE has been reported, angle dependent measurements are always required [33]. On the other hand, stray magnetic field is generated regardless of whether one has in-plane or out-of-plane magnetization, and hence NV measurement can readily probe magnetization of any given direction, making NV a versatile technique.

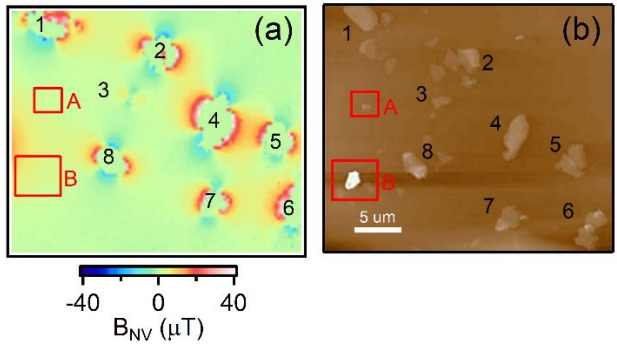

Figure 5. (a) The $B_{NV}$ stray field map of $Fe_5GeTe_2$ flakes without Pt protection layer. Stray field in areas where SNR is low is not shown. (b) The AFM image of the flakes in (a). The flake or the flakes cluster are labeled by numbers and the measured flakes thicknesses are #1(45~80 nm), #2(35~100 nm), #3(40~65 nm), #4(110 nm), #5(75~150 nm), #6(85 nm), #7(60 nm), #8(100 nm).

In addition, we demonstrate the utility of our QMI technique for rapid screening of magnetism in exfoliated flakes. Multiple $Fe_5GeTe_2$ flakes without Pt protection layer are imaged at room temperature and the stray field map is shown in figure 5(a). We note that in this measurement, $B_{NV}$ was imaged, hence the stray field pattern looks different from those in figure 2. As discussed previously, SNR of stray field measurement is low near the center of the flakes, so the stray field values are not displayed for these pixel points. The AFM image of the corresponding flakes is shown in figure 5(b). We have properly labeled the flakes or flake clusters in both panels. It is seen that some small flakes are visible in AFM picture (for example, flakes circled by red boxes A and B) but do not generate stray field. This may be attributed to the fact that those flakes are too small or too thin and hence have completely degraded in the absence of protective layer. However, even some larger flakes (for example, the flakes around the cluster #2) also do not generate stray field, which may also indicate complete degradation. Among the flakes that generate stray fields, we find the thinnest flake is 40 nm thick (< 20 unit cells). This result shows our imaging tool can measure the magnetism of unprotected $Fe_5GeTe_2$ flake also down to very thin layers. It also provides an option to investigate the degradation of the magnetic flakes in an ambient condition by imaging the sample field changing with time. Lastly, this work demonstrates the utility of QMI for rapid, parallel characterization of the thin flakes of other potential 2D magnets near room temperature, providing an enabling tool that will aid the effort in discovering additional 2D magnets that are useful for applications.

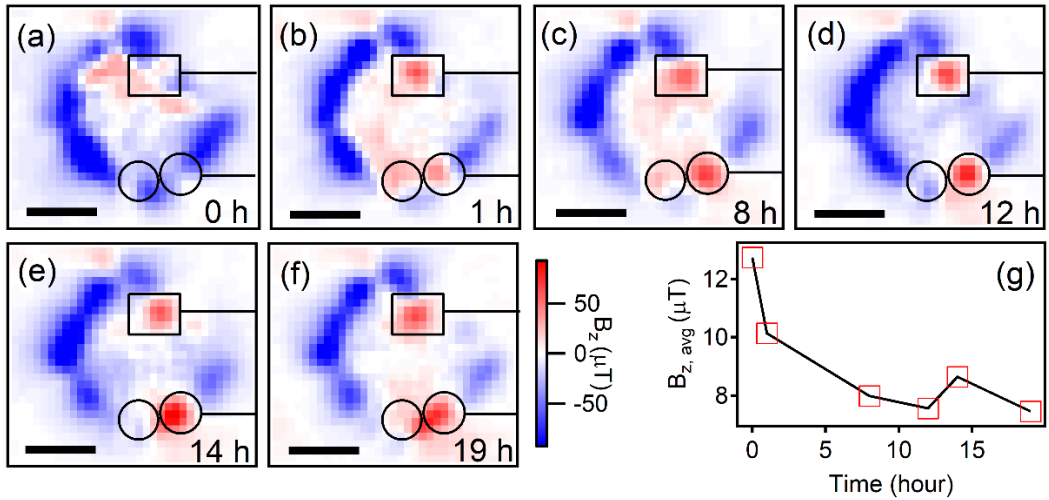

Figure 6. (a)-(f) Evolution of stray field image of the flake with exposure time in ambient condition. It is noted that here we use a color table that differs from previous color mappings to highlight the observed pining sites. Scale bar corresponds to 5 μm. The numbers indicate the evolution time *t* in hours in reference to the first image (a). The top pining site in each image is labeled with a black square and the bottom two pinning sites are labeled with two black circles. The positions of square and circles are fixed in each plot to help to identify the motion of pinning sites. (g) Absolute value of mean stray fields as a function of exposure time.

At last, we show the capability of QMI to monitor the time-dependent local magnetic properties of $Fe_5GeTe_2$ flakes from the evolution of its magnetic stray field with time *t*. This is motivated by the aspects of our results that point to additional local information that QMI can reveal compared to transport measurement. For instance, in figure 2(a,b), several orange spots around the flake edge indicate the existence of magnetic pinning in the flake; similar pinning behaviors have also been observed via scanning nitrogen vacancy magnetometry of another 2D magnet $CrBr_3$ at cryogenic temperature [21]. In figure 2(a,b), the orange spot is always paired with a blue spot, which indicates the local magnetic moment has a different orientation from the surrounding region. In addition, QMI can also reveal domain structures in the flake with terrace texture as shown in figure 2(e). This type of characterization is difficult to accomplish with electric transport measurement.

It is widely known that 2D magnetic materials are generally air-sensitive, and hence magnetic properties of the flake will possibly change with time due to oxidation of the material. Due to the lack of a suitable probe, there has been no report of magnetic degradation monitoring in the literature, except very recently in Ref. [34]. QMI's ability for parallel acquisition of stray field over a wide field of view enables monitoring of time-dependent local magnetic properties in response to degradation. In figure 6, we show the evolution of the stray fields with time by continuously imaging a flake (protected with a Pt layer) for twenty hours under the ambient condition. While the Pt layer provides an overall protection, one can anticipate certain degradation would occur due to imperfect insulation from the environment. Given the shorter amount of signal averaging in this measurement, we find SNR is poorer. Thus, we remove spurious pixels and apply pixels binning to better present the images and highlight the time-dependent behavior (see **S7** in Supplemental Material for more details). The flake has been initially magnetized in perpendicular direction and thus one observes a single-domain-like magnetic pattern corresponding to the expected stray field of out-of-plane magnetization as shown in figure 6(a). Within 1 hour of the initial measurement, three pinning sites (red pixels, labeled with square or circle) formed around the edge of the flake. It is found that pinning site prefers to appear near the edges of the flake, potentially due to imperfect covering of the protective layer at the edges.  This observation is consistent with the conclusion in Ref [34] where they attribute the formation of multiple magnetic domains in $Fe_3GeTe_2$ flake to oxygen absorption. Moreover, we also observe the movement of the formed domains with time. It is observed that the top pining site (labeled with square) is quite stable in the following twenty hours after it was formed in figure 6(b). However, the bottom-left pining site in figure 6(b) starts to move to the right. The stray field intensity of the bottom-right pinning site becomes stronger while the intensity of bottom-left one decreases with time until the blue pixels dominate in the left circled area as shown in figure 6(d). Finally, at $t = 14$ h, two bottom pining sites merge into a bigger one with a stronger intensify as seen in figure 6(e) and (f). In figure 6(g), we plot the average of the stray fields taken from each image as a function of time. It clearly shows that the overall stray fields have decreased by ~33% in the first ten hours and then settled into equilibrium state.  Ref. [34] reported a sharp drop in $M_s$~$t$ curve of an 8-layer $Fe_3GeTe_2$ flake within 30 minutes of exposure to atmosphere at 80 K [34]. Here, we monitor the magnetic change of the flake at room temperature for a much longer time, allowing us to observe the motion of the evolution of magnetic domains. These results provide critical insights into the evolution of local properties in 2D magnets due to oxidation.

## 4. Conclusion

In conclusion, we have employed QMI based on NV ensemble in diamond to study the magnetism of exfoliated $Fe_5GeTe_2$ thin flakes at room temperature. The measured stray fields have been confirmed to originate from the ferromagnetic moments in the flakes by the observation of different field intensities after magnetizing the flakes in opposite directions. The stray field patterns and response to magnetizing field point to the scenario that perpendicular easy axis anisotropy dominates in these flakes. Room-temperature ferromagnetism is observed in flake with thicknesses as low as 21 nm (7 units cell). The temperature dependent measurements allow us to determine the Curie temperature of the flakes $T_C \approx 300$ K which is higher than the room temperature and independent of the flake thickness in the range of 21 to 100 nm. These results pave the way towards realizing room-temperature monolayer ferromagnet with $Fe_5GeTe_2$. As a novel tool for characterizing microscopic magnetic materials and structures, our imaging technique provides a sensitive and artifact-free way to detect the minute signal of local magnetism. We also demonstrate the utility of our QMI setup in rapid screening of magnetism in exfoliated flakes, which enables the parallel measurement of many flakes at room temperature and provide a useful tool for characterization and screening of other potential near-room-temperature 2D magnets. At last, we show the utility of this tool in monitoring the evolution of magnetic property of the exfoliated flake in ambient condition. The time-dependent results demonstrate the motion of magnetic pinning sites which results in the decrease of the overall magnetization in the flake. This work paves the way for further study and characterization of few-layer room-temperature 2D magnets using NV quantum sensors, including the study of domains and spin textures [21,23], spin excitations [35], and control via spin transfer [36] or spin orbit torque [37]. The understanding of these aspects will lay the foundation for the application of such materials.

# References

[1] Z. Fei *et al*  2018 Two-dimensional itinerant ferromagnetism in atomically thin $Fe_3GeTe_2$ *Nat Mater* **17**  778

[2] J. Stahl, E. Shlaen and D. Johrendt  2018 The van der Waals Ferromagnets $Fe_{5-\delta}GeTe_2$ and $Fe_{5-\delta-x}Ni_xGeTe_2$ - Crystal Structure, Stacking Faults, and Magnetic Properties *Zeitschrift für anorganische und allgemeine Chemie* **644**  1923

[3] C. Tan *et al*  2018 Hard magnetic properties in nanoflake van der Waals $Fe_3GeTe_2$ *Nat Commun* **9**  1554

[4] M. Joe, U. Yang and C. Lee  2019 First-principles study of ferromagnetic metal $Fe_5GeTe_2$ *Nano Materials Science* **1**  299

[5] J. F. Sierra, J. Fabian, R. K. Kawakami, S. Roche and S. O. Valenzuela  2021 Van der Waals heterostructures for spintronics and opto-spintronics *Nat Nanotechnol* **16**  856

[6] C. Gong *et al*  2017 Discovery of intrinsic ferromagnetism in two-dimensional van der Waals crystals *Nature* **546**  265

[7] B. Huang *et al*  2017 Layer-dependent ferromagnetism in a van der Waals crystal down to the monolayer limit *Nature* **546**  270

[8] A. F. May, S. Calder, C. Cantoni, H. Cao and M. A. McGuire  2016 Magnetic structure and phase stability of the van der Waals bonded ferromagnet $Fe_{3-x}GeTe_2$ *Physical Review B* **93**  014411

[9] S. Liu *et al*  2017 Wafer-scale two-dimensional ferromagnetic $Fe_3GeTe_2$ thin films grown by molecular beam epitaxy *npj 2D Materials and Applications* **1**  30

[10] A. F. May *et al*  2019 Ferromagnetism Near Room Temperature in the Cleavable van der Waals Crystal $Fe_5GeTe_2$ *ACS Nano* **13**  4436

[11] A. F. May, C. A. Bridges and M. A. McGuire  2019 Physical properties and thermal stability of $Fe_{5-x}GeTe_2$ single crystals *Physical Review Materials* **3**  104401

[12] Z. Li *et al*  2020 Magnetic critical behavior of the van der Waals $Fe_5GeTe_2$ crystal with near room temperature ferromagnetism *Sci Rep* **10**  15345

[13] T. Ohta, K. Sakai, H. Taniguchi, B. Driesen, Y. Okada, K. Kobayashi and Y. Niimi  2020 Enhancement of coercive field in atomically-thin quenched $Fe_5GeTe_2$ *Applied Physics Express* **13**  043005

[14] A. F. May, M.-H. Du, V. R. Cooper and M. A. McGuire  2020 Tuning magnetic order in the van der Waals metal $Fe_5GeTe_2$ by cobalt substitution *Physical Review Materials* **4**  074008

[15] C. Tian, F. Pan, S. Xu, K. Ai, T. Xia and P. Cheng  2020 Tunable magnetic properties in van der Waals crystals $(Fe_{1-x}Co_x)_5GeTe_2$ *Applied Physics Letters* **116**  202402

[16] T. Ohta, M. Tokuda, S. Iwakiri, K. Sakai, B. Driesen, Y. Okada, K. Kobayashi and Y. Niimi  2021 Butterfly-shaped magnetoresistance in van der Waals ferromagnet $Fe_5GeTe_2$ *AIP Advances* **11**  025014

[17] F. Casola, T. van der Sar and A. Yacoby  2018 Probing condensed matter physics with magnetometry based on nitrogen-vacancy centres in diamond *Nature Reviews Materials* **3**  1

[18] M. J. H. Ku *et al*  2020 Imaging viscous flow of the Dirac fluid in graphene *Nature* **583**  537

[19] P. Maletinsky, S. Hong, M. S. Grinolds, B. Hausmann, M. D. Lukin, R. L. Walsworth, M. Loncar and A. Yacoby  2012 A robust scanning diamond sensor for nanoscale imaging with single nitrogen-vacancy centres *Nat Nanotechnol* **7**  320

[20] F. Fabre, A. Finco, A. Purbawati, A. Hadj-Azzem, N. Rougemaille, J. Coraux, I. Philip and V. Jacques  2021 Characterization of room-temperature in-plane magnetization in thin flakes of $CrTe_2$ with a single-spin magnetometer *Physical Review Materials* **5**  034008

[21] Q. C. Sun et al  2021 Magnetic domains and domain wall pinning in atomically thin $CrBr_3$ revealed by nanoscale imaging *Nat Commun* **12**  1989

[22] L. M. Pham et al  2011 Magnetic field imaging with nitrogen-vacancy ensembles *New Journal of Physics* **13**  045021

[23] D. A. Broadway et al  2020 Imaging Domain Reversal in an Ultrathin Van der Waals Ferromagnet *Adv Mater* **32**  e2003314

[24] J. F. Barry, M. J. Turner, J. M. Schloss, D. R. Glenn, Y. Song, M. D. Lukin, H. Park and R. L. Walsworth  2016 Optical magnetic detection of single-neuron action potentials using quantum defects in diamond *Proceedings of the National Academy of Sciences* **113**  14133

[25] J. P. Tetienne et al  2018 Spin properties of dense near-surface ensembles of nitrogen-vacancy centers in diamond *Physical Review B* **97**  085402

[26] J. P. Tetienne, D. A. Broadway, S. E. Lillie, N. Dontschuk, T. Teraji, L. T. Hall, A. Stacey, D. A. Simpson and L. C. L. Hollenberg  2018 Proximity-Induced Artefacts in Magnetic Imaging with Nitrogen-Vacancy Ensembles in Diamond *Sensors (Basel)* **18**  1290

[27] P. E. Gaskell, H. S. Skulason, C. Rodenchuk and T. Szkopek  2009 Counting graphene layers on glass via optical reflection microscopy *Applied Physics Letters* **94**  143101

[28] R. R. Nair, P. Blake, A. N. Grigorenko, K. S. Novoselov, T. J. Booth, T. Stauber, N. M. Peres and A. K. Geim  2008 Fine structure constant defines visual transparency of graphene *Science* **320**  1308

[29] V. L. Nguyen, D. L. Duong, S. H. Lee, J. Avila, G. Han, Y. M. Kim, M. C. Asensio, S. Y. Jeong and Y. H. Lee  2020 Layer-controlled single-crystalline graphene film with stacking order via Cu-Si alloy formation *Nat Nanotechnol* **15**  861

[30] Z. Ni, H. Wang, J. Kasim, H. Fan, T. Yu, Y. Wu, Y. Feng and Z. Shen  2007 Graphene thickness determination using reflection and contrast spectroscopy *Nano letters* **7**  2758

[31] Y. Liu, L. Wu, X. Tong, J. Li, J. Tao, Y. Zhu and C. Petrovic  2019 Thickness-dependent magnetic order in $CrI_3$ single crystals *Sci Rep* **9**  13599

[32] J. Seo et al  2020 Nearly room temperature ferromagnetism in a magnetic metal-rich van der Waals metal *Science advances* **6**  eaay8912

[33] A. Westphalen, M. S. Lee, A. Remhof and H. Zabel  2007 Invited article: Vector and Bragg Magneto-optical Kerr effect for the analysis of nanostructured magnetic arrays *Rev Sci Instrum* **78**  121301

[34] Z. Tu et al  2021 Ambient effect on the Curie temperatures and magnetic domains in metallic two-dimensional magnets *npj 2D Materials and Applications* **5**  62

[35] T. van der Sar, F. Casola, R. Walsworth and A. Yacoby  2015 Nanometre-scale probing of spin waves using single-electron spins *Nat Commun* **6**  7886

[36] A. Solyom, Z. Flansberry, M. A. Tschudin, N. Leitao, M. Pioro-Ladriere, J. C. Sankey and L. I. Childress  2018 Probing a Spin Transfer Controlled Magnetic Nanowire with a Single Nitrogen-Vacancy Spin in Bulk Diamond *Nano Lett* **18**  6494

[37] H. Zhang et al  2020 Spin-torque oscillation in a magnetic insulator probed by a single-spin sensor *Physical Review B* **102**  024404